\documentclass[11pt,letterpaper]{article}
\usepackage[margin=1in]{geometry}
\usepackage{times}
\usepackage[T1]{fontenc}
\usepackage{CJKutf8}
\makeatletter
\let\manuscriptCJKaddcmap\CJK@addcmap
\def\CJK@addcmap#1{%
  \manuscriptCJKaddcmap{#1}%
  \pdfnobuiltintounicode\font@name}
\makeatother
\usepackage[dvipsnames]{xcolor}
\usepackage{graphicx,amsmath,amssymb,booktabs,tabularx}
\usepackage[numbers,sort&compress]{natbib}
\usepackage[format=plain,labelformat=simple,labelsep=period,font=small,hypcap=false]{caption}
\usepackage[shortlabels,inline]{enumitem}
\usepackage[hyphens]{url}
\ExplSyntaxOn
\NewDocumentCommand{\tnum}{m}{
  \group_begin:
  \tl_set:Nn \l_tmpa_tl {#1}
  \tl_replace_all:Nnn \l_tmpa_tl {-} {\textminus}
  \textup{\l_tmpa_tl}
  \group_end:
}
\ExplSyntaxOff
\newcommand{\est}[2]{\mbox{\tnum{#1}\kern.03em\raisebox{-.24ex}{\normalfont\fontsize{6.6}{7.2}\selectfont\color{black!72}\tnum{#2}}}}

\newcolumntype{Y}{>{\raggedright\arraybackslash}X}
\DeclareUnicodeCharacter{2212}{\ensuremath{-}}
\definecolor{linkblue}{rgb}{0.21,0.49,0.74}
\usepackage[unicode,pagebackref,breaklinks,colorlinks,allcolors=linkblue,
  pdftitle={Equal Dose, Different Coverage: Tracing Rate and Duration in Image Editing},
  pdfauthor={Yi Liu}]{hyperref}

\title{Equal Dose, Different Coverage:\\Tracing Rate and Duration in Image Editing}
\author{Yi Liu\\
University of Science and Technology of China\\
{\tt\small scnuliuyi@mail.ustc.edu.cn}}
\date{}

\begin{document}
\maketitle
\par\noindent\begin{minipage}{\textwidth}
\centering
\includegraphics[width=\linewidth]{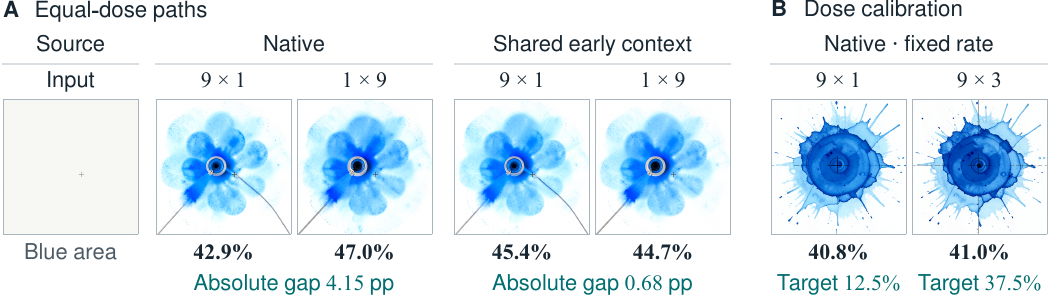}
\captionsetup{hypcap=false}
\captionof{figure}{\textbf{Route agreement and dose accuracy are separate requirements.} Examples use Chinese prompts with distilled FLUX. Column labels give rate $\times$ duration, and percentages give whole-canvas blue area. (A) The instruction fixes the deposition profile but prescribes no area scale, so the two nine-unit delivery routes should agree. Sharing representations at unchanged text positions (context) early in denoising narrows their gap in this layout from 4.15 to 0.68~pp. (B) With 36 units prescribed to cover 50\% of the canvas, the 9- and 27-unit targets are 12.5\% and 37.5\%. Native (unpatched) editing without size words leaves coverage nearly unchanged when duration triples. Table~\ref{tab:common-accuracy} tests sharing under this scale with matching size cues.}
\label{fig:teaser}
\end{minipage}\par
\vspace{1em}

\begin{abstract}
Benchmarks for scientific image editing test whether a model draws the outcome of a described process, leaving open which representations carry its errors and whether correcting them yields the right quantity. We study this question with an airbrush of fixed deposition profile, under which exchanging rate and duration preserves the dose and should preserve the endpoint, yet changes the edited coverage. Matched-norm activation patches across text positions and denoising stages show that, in distilled FLUX with Chinese prompts, representations at unchanged text positions (context) carry this discrepancy: early context transfers 3.51 of the 3.98 percentage-point (pp) signed route difference, against 0.09~pp for equal-norm updates at the two changed digits. Early image latents transmit native (unpatched) route differences across two FLUX configurations and OmniGen, while text-position effects depend on configuration. The trace predicts an intervention: share early context between routes while each retains its own digits. On independent English layouts, this reduces the distilled-FLUX absolute route gap from 1.13 to 0.60~pp, outperforms early digit sharing, and improves agreement more than in base FLUX under directly paired tests. Calibrated evaluation, which prescribes an area for each dose, separates this gain from accuracy: sharing the full prompt encoding produces identical route images with target errors of 8--16~pp, and stating the target percentage directly retains a compressed area response. Localization predicts a consistency gain on new layouts, and calibration measures the quantity error that agreement leaves unresolved.
\end{abstract}

\section{Introduction}
\label{sec:introduction}

Scientific image editing requires a model to translate a described process into its visible consequences. Filling a tank after a tap runs at a specified rate for a specified duration requires both inferring the accumulated volume and drawing the corresponding waterline. Science-T2I compares scientific process descriptions with explicit outcomes~\cite{li2025science}, and WorldEdit evaluates edits specified through their causes~\cite{lin2026worldedit}. Chain of Time reports similar predicted water-level slopes for image sequences with a threefold difference in flow rate~\cite{wang2025chain}. These evaluations score the drawn outcome, leaving open which representations transmit differences between process instructions, and whether equalizing those representations yields the correct quantity. We trace differences between equal-dose instructions, test the intervention this trace predicts on independent layouts, and measure its quantity accuracy.

One unit per second for nine seconds delivers as much water as nine units per second for one second. In a straight-sided 36-unit tank, both instructions require a quarter fill. Two half-full edits would agree and both be wrong. We therefore distinguish the gap between equivalent deliveries from each edit's distance to their common target. Our main task uses a fixed airbrush whose prescribed pigment profile is repeated for every delivered unit: equal doses deposit the same amount at every location. This deposition rule determines which outputs should agree. A calibrated version additionally states how dose maps to area, making absolute accuracy measurable (Fig.~\ref{fig:teaser}).

Exchanging rate and duration changes two digits in the instruction. Words that stay the same can acquire different representations in the two prompts. We call the two changed positions the digit group and the remaining valid, non-padding positions the context group. Following work that edits contextualized words to alter spatial relations~\cite{wang2026circuits}, we patch each group with the other route's embedding difference at equal norm to trace which one transmits the coverage difference. Generation stage determines when these representations act: early caption changes rearrange scenes, whereas later changes largely preserve their composition~\cite{balaji2022ediff}. Crossing the two groups with early and late denoising therefore locates where and when the route difference enters the final amount.

With Chinese prompts and distilled FLUX.2 [klein] 4B~\cite{bfl2026klein}, early context transfers 3.51 of the 3.98 percentage points (pp) of route difference. A common English protocol extends the trace to the undistilled base configuration and to OmniGen~\cite{xiao2025omnigen}. Early image states carry the native (unpatched) route difference in all three configurations, whereas text-position effects differ between them. The localization thus makes a testable prediction: on new layouts, sharing early context should reduce the gap more than sharing early digits, and more in distilled FLUX, whose context transfer is stronger. An independent English cohort tests this prediction for both $9\times1$ versus $1\times9$ and $4\times2$ versus $2\times4$.

Agreement alone leaves the common amount unconstrained. Scoring the same calibrated outputs by route gap and by target error shows that context sharing narrows gaps while substantial target error remains. These results separate a transferable discrepancy between equivalent processes from the error in rendering their common quantity.

Our contributions are as follows:
\begin{itemize}[label=\textbullet]
\item Equal-dose route differences are transmitted through early image states in all three tested configurations, including OmniGen's reversed (fast-over-long) difference. In distilled FLUX, early context rather than the changed digits transmits the difference, and this context-versus-digit stage effect exceeds that of base FLUX. Position-shuffled and row-count controls attribute it to the ordered context group as a whole.
\item In distilled FLUX, sharing early context reduces route disagreement on independent Chinese and English layouts while each route keeps its own digits. As the trace predicts, the English validation shows its advantage over early digit sharing and a greater gap reduction than in base FLUX under directly paired comparisons.
\item Route agreement and dose accuracy remain distinct after intervention. Shared conditioning reduces disagreement, including identical images under full-encoding sharing, while calibrated target error persists. Stating the target percentage directly still yields a compressed response, so the error persists without any product to compute.
\end{itemize}

\section{Related Work}
\label{sec:related}

\paragraph{Inferring and drawing a quantity.}
Instruction-based editing~\cite{brooks2023instruct} is commonly assessed by whether the requested change is made and surrounding content is preserved~\cite{ryu2025hatie}. Process instructions leave the change itself to be inferred. KRIS-Bench evaluates edits that require factual, conceptual or procedural knowledge~\cite{wu2025kris}. RISEBench asks what happens to a leaf under a magnifying glass in sunlight~\cite{zhao2025rise}. PhyEditBench includes video-derived intermediate states, and PhysicEdit learns transition priors from video~\cite{guo2026phyeditbench,zhao2026physics}. Numerical evaluations examine generated object counts~\cite{ghosh2023geneval}, edits~\cite{lassoued2026abacus}, visual arithmetic~\cite{wang2026arithmetic}, and fractions~\cite{kajic2024geckonum}. CountGen corrects object counts during denoising~\cite{binyamin2025count}. Continuous edit strength~\cite{zarei2026slider,parihar2026kontinuous}, raster geometry~\cite{xu2026paintbench}, and relative size~\cite{li2026genscale} provide complementary quantity readouts. Our accumulation rule supplies two linked constraints: equal-dose processes should agree, and a prescribed scale determines whether their common amount is correct.

\paragraph{Representation and stage interventions.}
Caption switching identifies concept-dependent windows of influence~\cite{gorgun2026temporal}. Cross-attention maps and spatial diffusion features transfer image structure across edits~\cite{hertz2023prompt,tumanyan2023plug}. Causal mediation and tracing localize behavior in language models~\cite{vig2020causal,meng2022locating} and visual knowledge in text encoders~\cite{basu2024localizing}. Layer ablations identify the layers vital for FLUX editing~\cite{avrahami2025stable}, and identified internal features steer demographic bias~\cite{shi2025dissecting}. Editing embeddings improves attribute binding~\cite{zhuang2024magnet} and changes spatial relations~\cite{wang2026circuits}. Text encoders compose prompts progressively across layers~\cite{toker2024diffusionlens}, lexical items retain distinct visual meanings after contextualization~\cite{kaplan2026flow}, and word meanings with position tags preserve aggregate text fidelity~\cite{spingarn2026text}. Padding tokens shape outputs in some architectures~\cite{toker2025padding}, and in FLUX.2 their activations transfer reference color and style between edits with different input images~\cite{ge2026binding}. Activation-patching studies relate the inferred role of a representation to the prompt contrast, intervention direction and evaluation metric~\cite{zhang2024patching,heimersheim2024patching}. Our position, norm and stage controls trace a continuous area difference with the input image held fixed, and the resulting localization specifies which representations to share on an independent layout cohort.

\paragraph{Equivalent outputs and accurate outputs.}
MetaLogic uses logically equivalent prompts to test whether generated scenes preserve entities and relations~\cite{shen2025metalogic}. Counterfactual evaluation distinguishes reconstruction under a null intervention from attainment of an attribute target~\cite{monteiro2023axiomatic,melistas2024counterfactual}. Some generated clocks print the requested time while their hands show 10:10~\cite{cazzaniga2026clocks}. Here equivalent products of rate and duration determine a shared endpoint. Measuring route gap and target error on the same intervened outputs identifies which constraint the intervention improves and which error remains.

\section{Measuring a cumulative edit}
\label{sec:method}
In the digital airbrush task, a fixed nozzle deposits paint at constant rate \(r\) for active duration \(t\), giving dose \(q=rt\). Each unit adds the same radial pigment profile at the same location, and the instruction requests the final stable state. Exchanging rate and duration preserves the final amount at every location. We call the higher-rate delivery the fast route and the longer delivery the long route. The settings \((9,1)\) and \((1,9)\) deliver nine units, \((4,2)\) and \((2,4)\) deliver eight, and \((9,3)\) and \((3,9)\) deliver 27.

\paragraph{Equality and accuracy constraints.}
The natural instruction requires equal endpoints for equal doses without assigning either an absolute area. More pigment can also darken already-blue pixels, so its coverage need not grow proportionally with dose. A calibrated instruction explicitly makes area proportional to dose, with 36 units covering half the canvas. Its targets are 12.5\% at nine units and 37.5\% at 27. In the accompanying straight-sided 36-unit tank task, these doses require 25\% and 75\% fill. Absolute accuracy is evaluated under these calibrated instructions.

\paragraph{Signed difference, gap and target error.}
For an equal-dose pair, let \(D_{i,o,q}=A_{i,o,q,L}-A_{i,o,q,F}\). Here \(A\) is blue coverage as a percentage of the measurement region, \(i\) indexes a paired source layout and generation seed, \(o\) is clause order, and \(L,F\) are the long and fast routes. The mean signed difference measures directional bias. Disagreement regardless of direction is
\begin{equation}
 G(q)=\mathbb E_{i,o}\bigl[|D_{i,o,q}|\bigr],
 \label{eq:gap}
\end{equation}
where \(\mathbb E\) denotes an equally weighted sample mean. Absolute differences are computed within route pairs before averaging. Calibrated outputs are also scored by their individual-image mean absolute error (MAE),
\begin{equation}
 E=\mathbb E_{i,o,q,s}\bigl[|A_{i,o,q,s}-A^\star(q)|\bigr],
 \label{eq:accuracy}
\end{equation}
where \(A^\star(q)\) is the prescribed target and \(s\in\{L,F\}\) denotes the route. The two calibrated doses receive equal weight. A constant-area spot has zero route gap at both doses while its MAE is at least 12.5~pp.

\paragraph{Behavioral checks of the remaining error.}
The response to dose is \(\bar A(27)-\bar A(9)\), with \(\bar A\) averaging routes, orders and layouts. Prompts describe the process, append its correct total, or directly specify the target percentage. The percentage supplies the desired measurement without requiring multiplication. Crossing dose with small/large spot wording, or low/high tank wording, separates the effect of numerical dose at fixed wording from the effect of a qualitative cue at fixed dose. An additional \(\{2,4\}\times\{2,4\}\) process grid tests other numerical combinations.

\section{Intervening along the generation trajectory}
\label{sec:intervention}
In reciprocal interventions, part of each route's conditioning is replaced with a representation from the other route. The supplying route is the donor and the receiving route is the recipient. Native edits retain their own conditioning throughout.

\paragraph{Representation groups and matched strength.}
Within each clause order, paired FLUX prompts have identical attention masks and differ at exactly two digit positions. These form the digit group \(\mathcal D\). The context group \(\mathcal C\) contains all other valid positions, including unchanged numerical constants, and excludes padding. Let \(\delta\) be the donor embedding minus the recipient embedding. A digit exchange adds \(\delta_{\mathcal D}\). The matched context update is
\begin{equation}
 \delta'_{\mathcal C}
 =\delta_{\mathcal C}\frac{\|\delta_{\mathcal D}\|_F}{\|\delta_{\mathcal C}\|_F},
 \label{eq:patch}
\end{equation}
where \(\|\cdot\|_F\) is the Frobenius norm. The two updates thus have equal norm, and the context update keeps the direction of the context difference. All other recipient rows stay unchanged. A row-count control retains the two context rows with the largest changes and matches their joint norm and temporal support to the digit update. A strength check scales both matched updates by a common factor \(\alpha\) (Supp.\ Sec.~\ref{app:strength}).

\paragraph{Generation stage and transfer.}
Updates act during the early or late half of the sampling schedule, or throughout sampling. At all other steps the recipient retains its own encoding. Signed transfer is
\begin{equation}
 T_p=\tfrac12\big[(A_F^p-A_F)+(A_L-A_L^p)\big],
 \label{eq:transfer}
\end{equation}
where \(p\) denotes the intervention and the superscript identifies a patched recipient's final area. Both terms orient reciprocal changes along the native long-minus-fast difference. Full-encoding exchange throughout sampling gives \(T_p=A_L-A_F\), including for a negative native difference. The position-by-stage contrast is \(J=(T_{\mathrm{ctx,early}}-T_{\mathrm{ctx,late}})-(T_{\mathrm{dig,early}}-T_{\mathrm{dig,late}})\). It compares the early-minus-late effect between the two representation groups. Donor-area error \(R_p\) averages \(|A_F^p-A_L|\) and \(|A_L^p-A_F|\) within each pair, then across pairs. Its native baseline is \(G\), so positive \(G-R_p\) measures movement closer to donor coverage regardless of native sign.

\paragraph{State and placement controls.}
In the midpoint state exchange, the two routes swap native image latents at the sampling midpoint, and each trajectory finishes under the recipient's prompt. The recipient thus inherits the donor's early state, so \(T_p\) in Eq.~\ref{eq:transfer} gives the early contribution and \(A_L-A_F-T_p\) gives the late contribution. Assigning each crossed trajectory instead to the route that produced its early state defines the late exchange from the same images. No text alignment is needed, so the exchange applies to all three configurations. Text controls permute context updates across valid rows while holding norm and stage fixed. The English spatial control permutes the native midpoint latent difference before adding it to the recipient. The Chinese spatial control instead permutes the midpoint change induced by an early context patch. Both preserve update norms up to rounding.

\paragraph{From localization to an independent prediction.}
If early context transmits the route difference, supplying its pairwise mean to both routes should reduce their gap on new layouts. Early context sharing averages the two vectors at every context position for the first half of sampling. Each route retains its digits, padding and complete late conditioning. Controls share digits early or context late. Full-encoding sharing throughout sampling gives identical conditioning in the calibrated control. Sharing is computed directly from the paired prompt encodings. The representation groups, stage division and averaging rule are fixed before independent validation. Natural-task validation measures the route gap, and calibrated validation measures gap and target MAE on the same outputs.

\section{Experiments}
\label{sec:results}
\newcommand{\fixedcell}[2]{\tnum{#1}\phantom{\est{}{#2}}}
\newcommand{\tablenote}[1]{\par\smallskip\noindent
  \parbox[t]{\linewidth}{\raggedright\footnotesize\textit{Note.}\enspace #1}\par}

\subsection{Experimental setup}
\label{sec:setup}
\begin{table}[htbp]
\centering
\caption{\textbf{Cohorts and their roles in tracing and evaluating equal-dose edits.}}
\label{tab:design}
\small
\setlength{\tabcolsep}{4pt}
\begin{tabularx}{\linewidth}{@{}p{0.19\linewidth}lp{0.22\linewidth}Yr@{}}
\toprule
Task & Language & Configurations & Role & Paired units \\
\midrule
Natural airbrush & Chinese & FLUX distilled & Representation and stage tracing & 12 \\
Natural airbrush & English & FLUX distilled/base, OmniGen & Representation and stage tracing & 8 \\
Natural airbrush & Chinese & FLUX distilled & Independent sharing validation & 12 \\
Natural airbrush & English & FLUX distilled/base & Independent sharing validation & 12 \\
Calibrated airbrush & Chinese & FLUX distilled & Gap and target MAE after sharing & 12 \\
Calibrated airbrush & English & FLUX distilled/base & Gap and target MAE after sharing & 8 \\
Calibrated spot and tank & Chinese / English & FLUX distilled / all three & Numerical and qualitative controls & 12 / 8 \\
\bottomrule
\end{tabularx}
\par\smallskip\noindent\parbox{\textwidth}{\footnotesize\textit{Note.}\enspace A unit pairs one source layout with one generation seed. Chinese natural and calibrated validation share their held-out cohort. English strength checks reuse the tracing cohort.}
\end{table}
The Chinese experiments use frozen distilled FLUX.2-klein-4B~\cite{bfl2026klein}. The English protocol compares this model, FLUX.2-klein-base-4B and OmniGen~\cite{xiao2025omnigen}. We used four steps with guidance 1, 25 steps with guidance 4, and 25 steps with text/image guidance 2.5/1.6, respectively. Early interventions end after step 2 or 12, with late interventions covering the remaining steps. Configuration comparisons thus span weights and samplers.

English natural-task validation (Table~\ref{tab:design}) applies the prompt templates, phases and sharing rules fixed during tracing to new paired layouts. The strength check reuses the English tracing cohort at nine units and scales matched updates by 0.25, 0.5 and 1. Native and unit-scale outputs are reused under the same implementation.

The \(512\times512\) spot canvases vary in mark location, background tone and grid. Blue pixels satisfy \(2B-R-G>0.08\times510\) in 8-bit RGB. Coverage uses the whole spot canvas or the original tank interior. Tank pixels additionally require \(B-R>18\). Thresholds 0.06 and 0.12 and source-anchored connected regions provide readout sensitivity checks. Conditions within a layout use the same source, initial noise and reference encoding. Layouts and seeds are paired across configurations, and comparisons use paired layout differences.

The independent English validation prespecifies three primary comparisons: distilled early context versus native, distilled early context versus early digits, and the paired configuration difference in the gain from early context sharing. Both orders and operand combinations are averaged within each layout, and the three comparisons receive a joint Holm correction. Confidence intervals resample complete layout--seed units. Supplement Secs.~\ref{app:prompt}--\ref{app:sampling} give prompts, translations, hardware, software and sampling details, and Secs.~\ref{app:chinese}--\ref{app:readout} report uncertainty, per-operand results and the complete controls.

\subsection{The route difference is transmitted through early states}
\label{sec:behavior}
\label{sec:transfer}
The fixed deposition rule requires equal-dose routes to reach the same endpoint. Native edits violate this constraint. The long route covers more area in the Chinese experiment and, at both English operand combinations, in both FLUX configurations, whereas OmniGen favors the fast route. The opposite signs make OmniGen a test of whether the same generation stage carries a native discrepancy of either direction.

\begin{figure}[htbp]
  \centering
  \includegraphics[width=\textwidth]{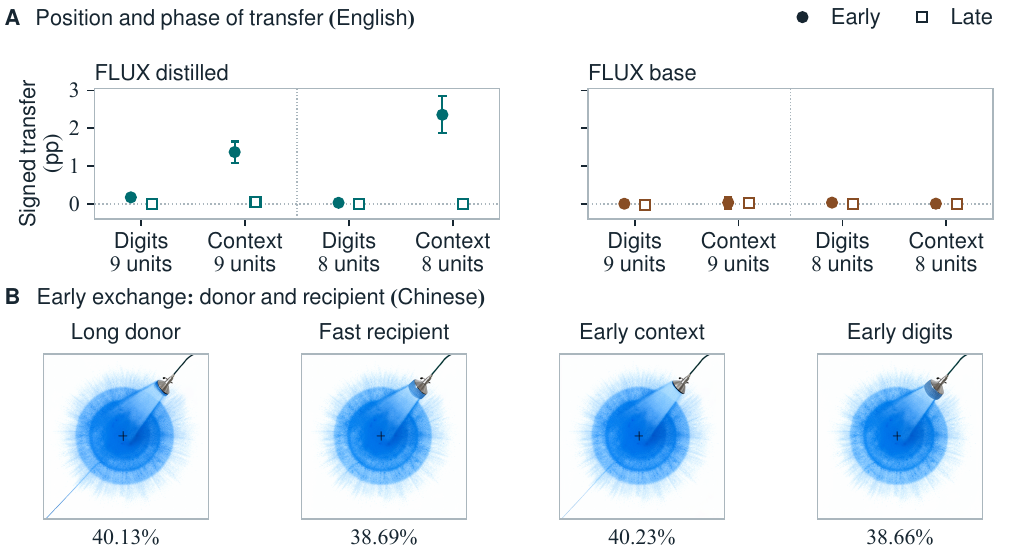}
  \caption{\textbf{Position and stage localize the route difference.} (A) Signed coverage transfer after norm-matched digit or context exchange. Points and bars give means and SEs. Early context carries a larger difference in distilled FLUX than in the base configuration. (B) Nine-unit routes: the long donor (1 unit/s for 9 s), fast native recipient (9 units/s for 1 s), and that recipient after early context or digit exchange. Context moves coverage toward the donor, while digits retain the recipient's coverage. Percentages measure whole-canvas blue area in the final outputs.}
  \label{fig:mechanism}
\end{figure}
The midpoint exchange tracks the native signed difference in all three configurations and at both operand combinations, including OmniGen's negative difference. Early exchange brings mean coverage closer to the donor than late exchange in all six comparisons. Spatially shuffling the midpoint difference increases donor-area error. These controls locate the transmission of route-dependent coverage in the spatial organization of early image states (Supp.\ Sec.~\ref{app:trace}).

Text interventions identify which conditioning representations transfer coverage. In the Chinese nine-unit experiment, early context transfers 3.51~pp of the 3.98~pp signed route difference, while equal-norm digit updates transfer 0.09~pp (Fig.~\ref{fig:mechanism}B). The position-by-stage contrast favors context on every layout. Shuffling the contextual updates across text positions, or spatially shuffling the latent change they induce, removes most of the transfer.

\begin{table}[htbp]
\centering
\caption{\textbf{Text transfer and donor-area comparisons.} Mean effects in percentage points of canvas area.}
\label{tab:common-text}
\small
\begin{tabular*}{\textwidth}{@{\extracolsep{\fill}}lrrrr@{}}
\toprule
 & \multicolumn{2}{c}{FLUX distilled} & \multicolumn{2}{c}{FLUX base} \\
\cmidrule(lr){2-3}\cmidrule(l){4-5}
Quantity / comparison & Nine units & Eight units & Nine units & Eight units \\
\midrule
Native absolute gap $G$ & \est{1.642}{0.328} & \est{1.822}{0.470} & \est{0.993}{0.420} & \est{1.180}{0.708} \\
\midrule
\multicolumn{5}{l}{\textit{Signed transfer $T$ and stage interaction $J$}} \\
Two digits, all steps & \est{+0.170}{0.054} & \est{+0.038}{0.048} & \est{-0.024}{0.061} & \est{+0.020}{0.026} \\
Context, matched norm, all steps & \est{+1.401}{0.283} & \est{+2.356}{0.493} & \est{+0.084}{0.153} & \est{+0.004}{0.108} \\
Context, shuffled rows, all steps & \est{+0.281}{0.506} & \est{-0.236}{0.286} & \est{+1.182}{1.181} & \est{+0.338}{0.392} \\
Two context rows, all steps & \est{+0.190}{0.067} & \est{-0.078}{0.045} & \est{+0.175}{0.121} & \est{+0.070}{0.021} \\
Position-by-stage interaction & \est{+1.147}{0.231} & \est{+2.333}{0.496} & \est{-0.037}{0.098} & \est{-0.037}{0.100} \\
\midrule
\multicolumn{5}{l}{\textit{Donor-area advantage: comparator $R$ minus treatment $R$}} \\
Context versus digits & \est{+0.895}{0.191} & \est{+0.997}{0.440} & \est{+0.027}{0.156} & \est{+0.003}{0.082} \\
Ordered versus shuffled context & \est{+1.226}{0.396} & \est{+1.452}{0.631} & \est{+0.345}{0.107} & \est{-0.329}{0.382} \\
Two context rows versus digits & \est{+0.026}{0.052} & \est{-0.108}{0.065} & \est{-0.117}{0.169} & \est{-0.016}{0.023} \\
\bottomrule
\end{tabular*}
\par\smallskip\noindent\parbox{\textwidth}{\footnotesize\textit{Note.}\enspace Subscripts give standard errors. For the four $T$ rows and all donor-error comparisons, updates act throughout sampling. $J$ compares the early-minus-late transfer of context with that of digits. Positive donor-error differences indicate lower donor-area error for the first named treatment. Donor-error comparisons are post hoc sensitivity analyses.}
\end{table}
Under the common English protocol, distilled FLUX has a larger context-versus-digit position-by-stage effect than base FLUX at both operand combinations (Table~\ref{tab:common-text}). All-step context exchange also produces a greater donor-area advantage over digits (lower \(R_p\)) in distilled FLUX. Its ordered context updates reproduce donor coverage more closely than shuffled updates. Retaining just the two context rows with the largest changes does not reproduce the full group's advantage over digits, so the advantage belongs to the context group as a whole rather than to its most-changed rows. The two FLUX configurations receive identical complete prompt encodings, source images and initial noise, locating their differing text effects downstream of the text representation. The larger interaction at eight than nine units further shows dependence on the operand combination, which changes both dose and factor ratio.

At \(\alpha=0.25\), 0.5 and 1, distilled FLUX retains a larger position-by-stage interaction than base FLUX and lower early-context than early-digit donor error. Base FLUX shows a positive interaction at \(\alpha=0.5\), yet its mean early-context donor error exceeds the native baseline at every strength. Larger signed transfer therefore does not imply output closer to the donor (Supp.\ Sec.~\ref{app:strength}).

\subsection{The trace predicts an intervention on independent layouts}
\label{sec:heldout}
Early context sharing tests the localization's prediction that equalizing the transmitting representations shrinks the route gap while digit vectors stay distinct. In the independent Chinese cohort, sharing early context reduces the nine-unit gap from 4.33 to 0.98~pp and also narrows the 27-unit gap. Early digit and late context sharing retain greater disagreement, matching the position and stage isolated by exchange.

\begin{figure}[htbp]
  \centering
  \includegraphics[width=\textwidth]{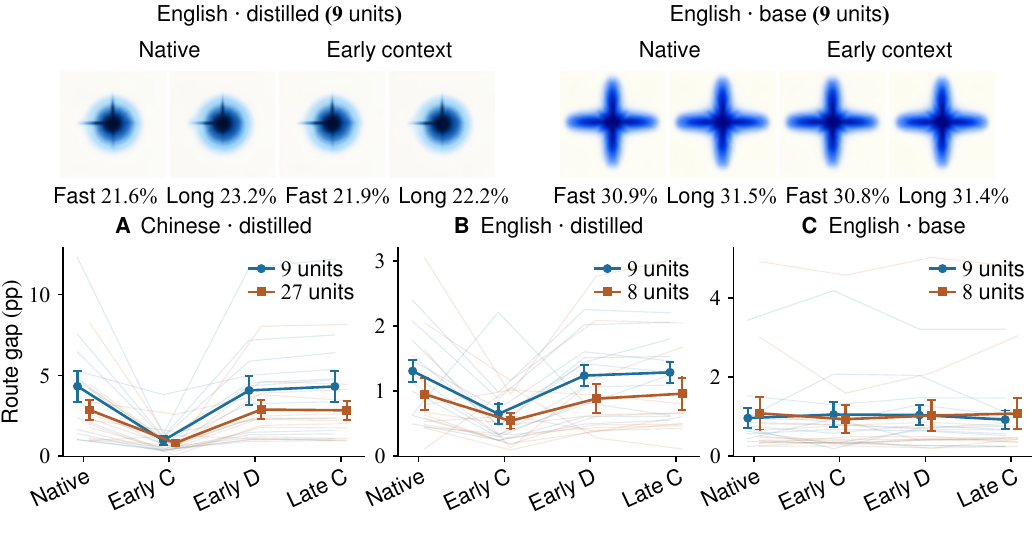}
  \caption{\textbf{Independent validation of the localization-derived intervention.} Top: English nine-unit fast and long routes before and after early context sharing in both configurations. Percentages give whole-canvas blue area. Bottom: absolute route gaps for every layout (thin paired lines), cohort means (markers) and SEs. C and D denote context and digits. Chinese and English panels retain their respective operand combinations.}
  \label{fig:independent}
\end{figure}
The independent English cohort confirms both the position prediction and its configuration dependence (Fig.~\ref{fig:independent}). In distilled FLUX, early context sharing reduces the gap, averaged over both operand combinations, from 1.13 to 0.60~pp and outperforms early digit sharing by 0.47~pp. The gap reduction exceeds that in base FLUX by 0.51~pp. All three prespecified comparisons remain significant after joint Holm correction (each $p_H=0.0264$). Late context sharing leaves the distilled gap at its native 1.13~pp, as the stage localization predicts. The distilled mean gap decreases at both operand combinations, whereas the base mean gap increases at nine units and decreases at eight, for a net reduction of 0.03~pp. All three primary directions hold at both alternative blue thresholds (Supp.\ Sec.~\ref{app:independent}).

\subsection{Agreement improves without calibrating the shared amount}
\label{sec:calibrated}
The calibrated task adds a common target and scores gap and MAE on the same outputs, with a small-spot cue at nine units and a large-spot cue at 27. Early context sharing narrows the Chinese gaps at both doses while MAE stays near 16~pp. In the English calibrated cohort, it narrows the nine-unit gap, while the 27-unit gap changes little in distilled FLUX and widens in base FLUX. MAE remains 13.35 and 8.36~pp, respectively (Table~\ref{tab:common-accuracy}).

\begin{table}[htbp]
\centering
\caption{\textbf{Route agreement and target accuracy on the same calibrated edits.} Absolute route gap $G$ and mean absolute target error (MAE), in percentage points of canvas area. The nine-/27-unit targets use small-/large-spot cues.}
\label{tab:common-accuracy}
\small
\begin{tabular*}{\textwidth}{@{\extracolsep{\fill}}lrrrrrrrrr@{}}
\toprule
 & \multicolumn{3}{c}{Chinese FLUX} & \multicolumn{3}{c}{English FLUX} & \multicolumn{3}{c}{English FLUX base} \\
\cmidrule(lr){2-4}\cmidrule(lr){5-7}\cmidrule(l){8-10}
Conditioning & $G(9)$ & $G(27)$ & MAE & $G(9)$ & $G(27)$ & MAE & $G(9)$ & $G(27)$ & MAE \\
\midrule
Native & \est{1.79}{0.76} & \est{0.48}{0.10} & \est{16.03}{0.91} & \est{0.16}{0.06} & \est{0.16}{0.02} & \est{13.35}{0.59} & \est{0.23}{0.11} & \est{0.82}{0.19} & \est{8.33}{1.45} \\
Share digits early & \est{1.61}{0.72} & \est{0.50}{0.10} & \est{16.01}{0.89} & \est{0.12}{0.03} & \est{0.19}{0.03} & \est{13.35}{0.59} & \est{0.26}{0.11} & \est{0.80}{0.17} & \est{8.34}{1.44} \\
Share context late & \est{1.79}{0.76} & \est{0.49}{0.10} & \est{16.03}{0.91} & \est{0.16}{0.05} & \est{0.16}{0.02} & \est{13.35}{0.59} & \est{0.22}{0.10} & \est{0.84}{0.19} & \est{8.33}{1.45} \\
Share context early & \est{1.02}{0.50} & \est{0.32}{0.05} & \est{16.02}{0.94} & \est{0.09}{0.03} & \est{0.17}{0.07} & \est{13.35}{0.59} & \est{0.10}{0.04} & \est{0.96}{0.25} & \est{8.36}{1.45} \\
Share full encoding & \fixedcell{0.00}{0.00} & \fixedcell{0.00}{0.00} & \est{15.96}{0.88} & \est{0.00}{0.00} & \est{0.00}{0.00} & \est{13.37}{0.60} & \est{0.00}{0.00} & \est{0.00}{0.00} & \est{8.49}{1.47} \\
\bottomrule
\end{tabular*}
\par\smallskip\noindent\parbox{\textwidth}{\footnotesize\textit{Note.}\enspace Subscripts give standard errors. MAE weights both doses equally. Full-encoding sharing is a technical control with identical route images and zero gaps.}
\end{table}
\begin{figure}[htbp]
  \centering
  \includegraphics[width=\textwidth]{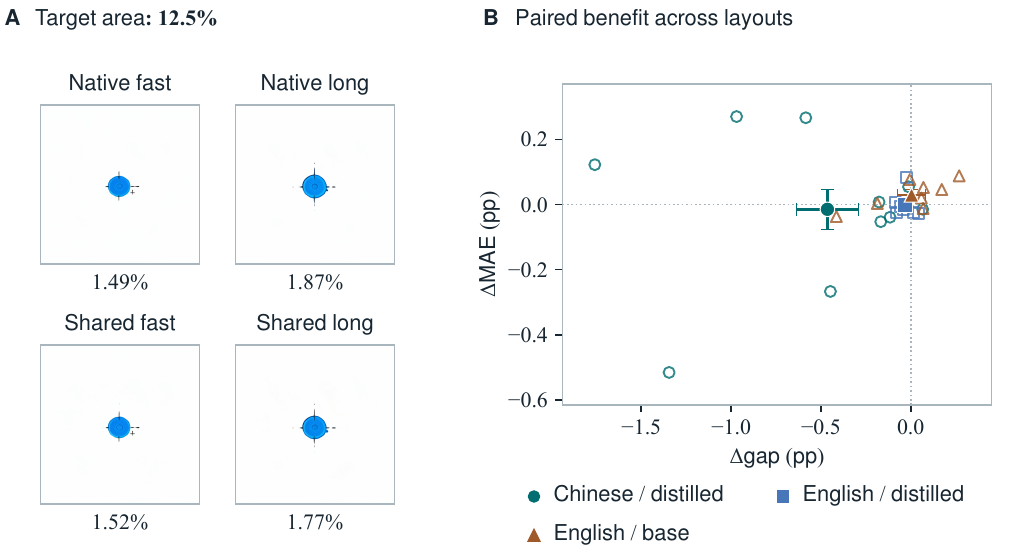}
  \caption{\textbf{Consistency and accuracy on the same calibrated edits.} (A) Native and early-context-shared Chinese outputs at a common nine-unit target of 12.5\% canvas area with the small-spot cue. Readouts show measured coverage. (B) Each open symbol gives one layout's change under early context sharing, averaged over orders and doses. Changes are shared minus native, so negative values improve the metric. Filled symbols and bars give cohort means and SEs. Route-gap reductions leave the shared quantity inaccurately calibrated.}
  \label{fig:target-errors}
\end{figure}
Full-encoding sharing produces identical route images, yet MAE remains 13.37~pp in English distilled FLUX and 8.49~pp in base FLUX. Figure~\ref{fig:target-errors} pairs the changes in gap and target error on each layout: improved agreement leaves the common quantity inaccurate.

For a pair with areas \(A_L,A_F\) and target \(a\), let \(m=(A_L+A_F)/2\) be the midpoint and \(d=|A_L-A_F|/2\) its half-gap. Then
\begin{equation}
 \frac{|A_L-a|+|A_F-a|}{2}=\max\{|m-a|,d\}.
 \label{eq:pair-error}
\end{equation}
Most native calibrated pairs miss on the same side of the target: 94\% in Chinese, all English distilled pairs and 94\% of English base pairs. For these pairs, Eq.~\ref{eq:pair-error} reduces to \(|m-a|\), the error in their common amount. As a numerical counterfactual, replacing both areas in each pair by their midpoint eliminates every route gap and lowers MAE by 0.16, 0 and 0.02~pp, respectively. Both this decomposition and the shared-conditioning outputs show that the dominant target error lies in the common amount.

\paragraph{Supplying the target leaves the response compressed.}
\label{sec:cuecross}
Explicit endpoints test whether compression persists when the desired amount is supplied (Table~\ref{tab:common-cuecross}). These checks return to native conditioning, and the process-only, supplied-total and direct-percentage prompts omit qualitative size cues. Tripling dose under process-only instructions moves English coverage by at most 0.49~pp in any configuration, against prescribed gains of 25~pp for the spot and 50~pp for the tank. With the correct total appended or the target percentage stated directly, every mean coverage change remains within 5~pp of zero. The direct-percentage control exposes an error in converting an explicit numerical target into visual area.

\begin{table}[htbp]
\centering
\caption{\textbf{Dose and qualitative appearance controls.} Mean paired changes in blue coverage over the whole canvas (spot) or the original tank interior, in percentage points.}
\label{tab:common-cuecross}
\footnotesize
\setlength{\tabcolsep}{2.5pt}
\begin{tabular*}{\textwidth}{@{\extracolsep{\fill}}lrrrrrrrr@{}}
\toprule
 & \multicolumn{2}{c}{Chinese FLUX} & \multicolumn{2}{c}{English FLUX} & \multicolumn{2}{c}{English FLUX base} & \multicolumn{2}{c}{English OmniGen} \\
\cmidrule(lr){2-3}\cmidrule(lr){4-5}\cmidrule(lr){6-7}\cmidrule(l){8-9}
Instruction contrast & Spot & Tank & Spot & Tank & Spot & Tank & Spot & Tank \\
\midrule
Prescribed gain, 9 to 27 units & \fixedcell{+25.00}{0.00} & \fixedcell{+50.00}{0.00} & \fixedcell{+25.00}{0.00} & \fixedcell{+50.00}{0.00} & \fixedcell{+25.00}{0.00} & \fixedcell{+50.00}{0.00} & \fixedcell{+25.00}{0.00} & \fixedcell{+50.00}{0.00} \\
Dose from process alone & \est{-0.27}{0.25} & \est{-0.21}{0.06} & \est{-0.09}{0.10} & \est{-0.08}{0.93} & \est{-0.08}{0.18} & \est{-0.49}{0.33} & \est{+0.02}{0.03} & \est{-0.05}{0.08} \\
Dose with correct total supplied & \est{+1.16}{0.42} & \est{+0.21}{0.06} & \est{+0.25}{0.10} & \est{-0.56}{0.50} & \est{+0.49}{0.36} & \est{+0.30}{0.51} & \est{-0.27}{0.23} & \est{+0.55}{0.34} \\
Target percentage alone & \est{+1.11}{0.31} & \est{-0.14}{0.23} & \est{+4.95}{2.05} & \est{+0.64}{0.43} & \est{+2.25}{1.40} & \est{+0.14}{0.08} & \est{-1.04}{0.48} & \est{-1.31}{0.98} \\
Dose at fixed size cue & \est{+1.26}{0.94} & \est{-0.04}{0.16} & \est{-0.63}{0.29} & \est{-1.16}{0.47} & \est{-0.18}{0.10} & \est{-0.01}{0.76} & \est{+0.01}{0.14} & \est{+0.42}{0.49} \\
Size cue at fixed dose & \est{+44.71}{4.07} & \est{+17.21}{3.16} & \est{+49.55}{3.12} & \est{+14.94}{5.95} & \est{+31.88}{8.11} & \est{+18.38}{6.95} & \est{+3.53}{3.10} & \est{+3.31}{1.30} \\
Dose and size cue together & \est{+45.97}{3.86} & \est{+17.17}{3.16} & \est{+48.92}{3.29} & \est{+13.78}{5.88} & \est{+31.70}{8.15} & \est{+18.38}{7.47} & \est{+3.54}{3.18} & \est{+3.73}{1.67} \\
\bottomrule
\end{tabular*}
\par\smallskip\noindent\parbox{\textwidth}{\footnotesize\textit{Note.}\enspace Subscripts give standard errors across layouts. FLUX denotes the distilled configuration. Chinese and English columns use separate cohorts. Prescribed gains are fixed targets.}
\end{table}
At fixed numerical dose, size words change spot coverage by 32--50~pp in both FLUX configurations, far beyond the effect of dose at fixed wording. The models respond to the requested appearance while failing to match its numerical scale. Compression also persists on the additional four-, eight- and 16-unit grid and under the source-anchored readout (Supp.\ Secs.~\ref{app:readout}--\ref{app:dose}).

\section{Discussion}
\label{sec:discussion}
Tracing route differences to early context identifies an effective sharing intervention. Its gains on independent layouts, and their dependence on configuration, establish the predictive value of the localization. Calibrated outputs then separate agreement from quantity accuracy: shared conditioning brings the routes together around an inaccurate common amount. The direct-percentage control shows that supplied numerical targets are inaccurately realized as visual area. Cumulative edits therefore require both equivalent-process consistency and quantity accuracy. Early context sharing improves consistency, while calibrating the common amount remains a distinct objective.

\section{Conclusion}
Equal doses produce different visual amounts in image editing. Position and stage interventions trace these differences through early image states and to context representations whose sharing improves agreement on independent layouts, more in distilled than in base FLUX. Calibrated targets show that improved agreement leaves substantial quantity error, so cumulative edits require both equivalent-process consistency and quantity accuracy.

{
  \small
  \bibliographystyle{plainnat}
  \bibliography{references}
}
\clearpage
\appendix
\setcounter{table}{0}\setcounter{figure}{0}
\renewcommand{\thetable}{S\arabic{table}}
\renewcommand{\thefigure}{S\arabic{figure}}
\renewcommand{\theHtable}{appendix.\arabic{table}}
\renewcommand{\theHfigure}{appendix.\arabic{figure}}
\begin{center}{\Large\bfseries Appendix\par}\end{center}
\renewcommand{\thesection}{S\arabic{section}}

\section{Prompt wording and translations}\label{app:prompt}
The Chinese prompt and its translation illustrate the nine-unit fast route. The English model input below gives the eight-unit fast route.

\paragraph{Chinese natural instruction.}
\begin{CJK*}{UTF8}{gbsn}编辑这张数字画布。在图上十字处固定一支圆形软喷枪，位置和喷幅设置始终不变。每单位蓝色颜料按相同的中心浓、边缘淡的圆形分布叠加。以每秒9个颜料单位的恒定速率喷涂蓝色颜料，持续1秒。画出喷涂停止后的最终稳定画布，保留原构图。\end{CJK*}

\paragraph{English translation.}
Edit this digital canvas. Fix a soft, round airbrush at the cross in the image. Its position and spray-width setting stay fixed throughout. Each unit of blue paint accumulates in the same circular pattern, dark at the center and light at the edges. Spray blue paint at a constant rate of 9 paint units per second. Continue for 1 second. Show the final stable canvas after spraying stops, preserving the original composition.

\paragraph{English model input.}
The English natural task uses the translated opening and final-state sentence above. Its two delivery sentences for the eight-unit fast route are “The constant delivery rate, in paint units per second, is 4. The active spraying duration, in seconds, is 2.”

\paragraph{Prompt variants.}
The reciprocal route swaps the two digits; duration-first wording reverses the clauses. OmniGen includes an image-reference prefix. The calibrated Chinese scale is “\begin{CJK*}{UTF8}{gbsn}蓝斑面积与累计量成正比，36单位对应整幅画布的一半。\end{CJK*}” Its English equivalent reads “The blue spot's area is proportional to the cumulative amount. Half of the entire canvas corresponds to 36 units.” Sharing pairs nine units with a small-spot cue and 27 with a large-spot cue. The behavioral task crosses cue with dose. For tanks, percentage specifies fill height: “The water level occupies 75\% of the full scale.”

\section{Sampling and experimental settings}\label{app:sampling}
\paragraph{Sampling and uncertainty.}
Within each layout, conditions share source and initial noise. Validation weights clause orders and operand pairs equally. Intervals use 20,000 whole-layout bootstrap samples. For the three primary English sharing contrasts (main-paper Table~1), Holm adjustment covers all three exact two-sided paired sign-flip tests. Standard errors use the between-layout standard deviation.

\paragraph{Hardware and software.}
Inference uses an NVIDIA RTX 4090 (24 GB), Python 3.12.3, PyTorch 2.8.0 (CUDA 12.8), Diffusers 0.42.1, Transformers 5.22.1 and Accelerate 1.15.0. All models use bfloat16 and flow-matching Euler sampling. FLUX uses dynamic exponential time shifting. OmniGen uses inverted sigmas and unit shift.

\section{Longer spraying increases coverage in the Chinese confirmation}\label{app:chinese}
At 90 units (Chinese, fixed canvas), 3 units/s for 30~s covered more area than 30 units/s for 3~s in 22 of 24 new seeds (mean gain 0.97~pp).

In Chinese distilled FLUX the position-by-stage contrast is 3.42~pp (95\% CI 2.09 to 4.99). Shuffling early context across text positions costs 3.85~pp of signed transfer (2.15 to 5.63). At matched norm, the two most-changed context rows exceed the two digit rows by 0.28~pp ($-0.03$ to 0.53).

\section{Early image states carry the route difference}\label{app:trace}
Across all six English model--dose comparisons, early state exchange yields lower donor-area error than late exchange. Spatial shuffling raises error even in OmniGen, which favors the fast route (Table~\ref{tab:state}).

\begin{table}[!htbp]
\centering
\caption{Decrease in donor-area error (percentage points, pp) under early versus late state exchange and intact versus spatially shuffled midpoint updates. Positive values favor the first condition.}
\label{tab:state}
\small
\begin{tabular*}{\linewidth}{@{\extracolsep{\fill}}lrrrr@{}}
\toprule
& \multicolumn{2}{c}{Early vs.\ late} & \multicolumn{2}{c}{Intact vs.\ shuffled}\\
\cmidrule(lr){2-3}\cmidrule(l){4-5}
& 9 units & 8 units & 9 units & 8 units\\
\midrule
FLUX distilled & 1.481 & 1.682 & 1.268 & 1.423\\
FLUX base      & 0.815 & 0.955 & 0.660 & 0.843\\
OmniGen        & 0.979 & 1.031 & 0.579 & 0.623\\
\bottomrule
\end{tabular*}
\end{table}
\section{Early context sharing narrows the gap on new layouts in distilled FLUX}
\label{app:independent}
On independent English layouts, early context gives the smallest distilled gap at both doses (Table~\ref{tab:validation}). In base FLUX its effect changes direction between doses.

\begin{table}[!htbp]
\centering
\caption{Independent English validation. Upper rows give mean route gaps in pp. Lower rows give prespecified paired gap reductions in pp and their 95\% bootstrap limits. C and D denote context and digits.}
\label{tab:validation}
\small
\begin{tabular*}{\linewidth}{@{\extracolsep{\fill}}lrrrr@{}}
\toprule
Model / dose & Native & Early C & Early D & Late C\\
\midrule
Distilled / 9 & 1.311 & 0.651 & 1.241 & 1.290\\
Distilled / 8 & 0.954 & 0.543 & 0.886 & 0.962\\
Base / 9      & 0.964 & 1.054 & 1.041 & 0.926\\
Base / 8      & 1.084 & 0.936 & 1.027 & 1.080\\
\bottomrule
\end{tabular*}
\smallskip
\begin{tabular*}{\linewidth}{@{\extracolsep{\fill}}lrrr@{}}
\toprule
Gap contrast & Mean & Lower & Upper\\
\midrule
Native $-$ early C, distilled & 0.535 & 0.206 & 0.860\\
Early D $-$ early C, distilled & 0.466 & 0.188 & 0.743\\
C benefit, distilled $-$ base & 0.507 & 0.171 & 0.845\\
\bottomrule
\end{tabular*}
\end{table}

All three primary contrasts give Holm-adjusted $p=0.0264$. Across blue thresholds 0.06 and 0.12, the distilled advantage over native is 0.47--0.59~pp and its paired advantage over base is 0.43--0.57~pp.

\section{Stronger early context updates reduce donor error in distilled FLUX}
\label{app:strength}
\begin{figure}[!htbp]
\centering
\includegraphics[width=\textwidth]{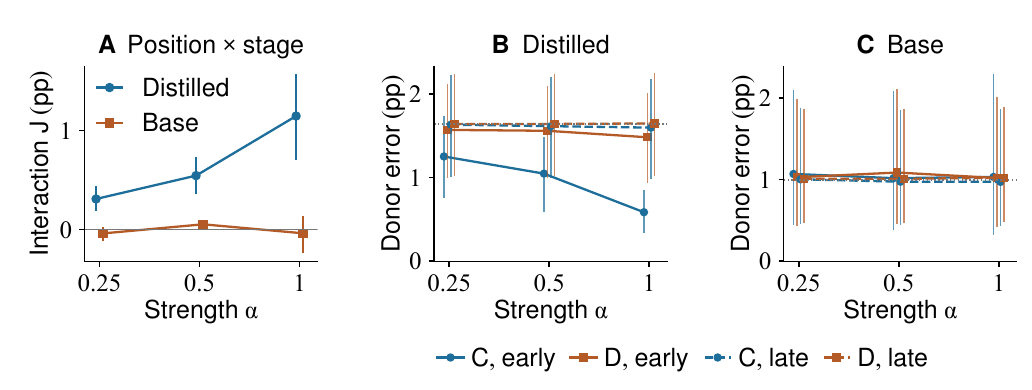}
\caption{\textbf{Sensitivity to intervention strength at nine units.} (A) The early-minus-late transfer advantage of context over digits ($J$) for each FLUX configuration. (B--C) Mean absolute area error relative to the donor after early or late exchange of context (C) or digits (D). Lower error indicates a closer match to the donor. Dotted lines give the native errors and bars give 95\% intervals from resampling paired layouts. Each tick doubles the matched update strength $\alpha$.}
\label{fig:strength}
\end{figure}
Digit and context updates at nine units are $\alpha\delta_{\mathcal D}$ and $\alpha\delta_{\mathcal C}/\rho$, with $\rho=\|\delta_{\mathcal C}\|_F/\|\delta_{\mathcal D}\|_F$. The shared encodings give $\rho=0.9776$ and $0.7950$ for rate-first and duration-first wording. Averaging applies half the context update ($\alpha=0.4888$ and $0.3975$).

Increasing strength raises $J$ and lowers early-context donor error in distilled FLUX (Fig.~\ref{fig:strength}). In base FLUX, $J=0.054$~pp at intermediate strength (95\% CI 0.015 to 0.087); donor error exceeds native at all strengths. Table~\ref{tab:strength-paired} compares configurations.

\begin{table}[!htbp]
\centering
\caption{Distilled minus base at each matched strength. $\Delta J$ compares position-by-stage interactions. $\Delta U$ compares the early-context advantage over early digits, $U=R_{\mathrm{digits,early}}-R_{\mathrm{context,early}}$. Values and paired-layout 95\% limits are in pp.}
\label{tab:strength-paired}
\small
\begin{tabular*}{\linewidth}{@{\extracolsep{\fill}}lrrrr@{}}
\toprule
$\alpha$ & $\Delta J$ & 95\% limits & $\Delta U$ & 95\% limits\\
\midrule
0.25 & 0.348 & 0.221 to 0.468 & 0.358 & 0.213 to 0.495\\
0.50 & 0.492 & 0.311 to 0.658 & 0.443 & 0.237 to 0.648\\
1.00 & 1.184 & 0.724 to 1.592 & 0.908 & 0.474 to 1.335\\
\bottomrule
\end{tabular*}
\end{table}
\section{Both routes usually miss the target on the same side}
\label{app:accuracy}
\begin{figure}[!htbp]
\centering
\includegraphics[width=\textwidth]{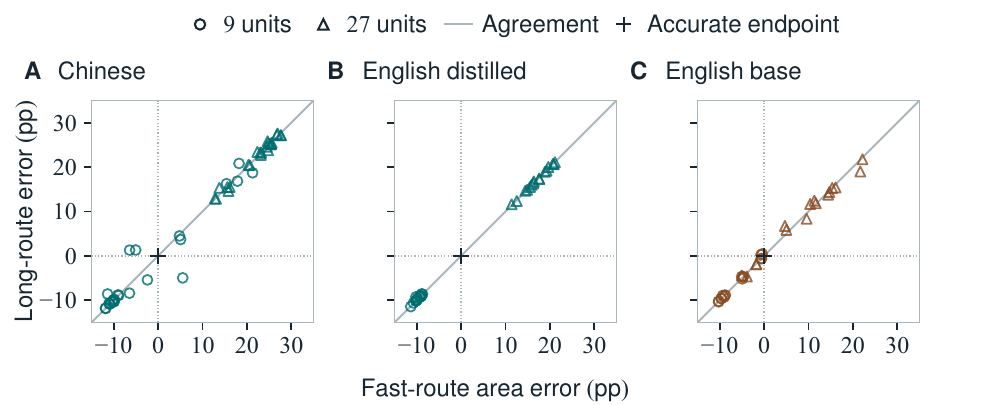}
\caption{\textbf{Common target error in native edits.} Each point pairs coverage errors at one dose. The diagonal marks route agreement and the origin marks target accuracy.}
\label{fig:native-errors}
\end{figure}

At all three thresholds, both routes miss on the same side in 45 of 48 native Chinese pairs (Fig.~\ref{fig:native-errors}). Early sharing changes Chinese dose-balanced MAE by $-0.015$~pp (95\% CI $-0.14$ to 0.09). At nine and 27 units, English distilled MAE changes from 9.62 to 9.63 and 17.08 to 17.06~pp, respectively. Base MAE changes from 6.07 to 6.08 and 10.59 to 10.64~pp.

\section{Coverage effects persist across blue thresholds}\label{app:readout}
The blue threshold $\tau$ sets the cutoff $2B-R-G>510\tau$. At $\tau=0.06$, 0.08 and 0.12, the nine-unit distilled-FLUX position-by-stage contrast is 1.33, 1.15 and 0.81~pp, respectively. The corresponding early-minus-late image-state contrasts are 1.76, 1.50 and 1.02~pp.

\section{Time, labels and quantitative edits}\label{app:scope}
\paragraph{Longer spraying covers more area under both wordings.}
Motivated by human duration--number interactions~\cite{dormal2006stroop,petrizzo2023topdown}, we compare seconds with event counts. Rephrasing reverses the Chinese event-route preference. Longer spraying covers more area in both wordings. The seconds-minus-events contrast shifts from $-2.88$ to 3.85~pp. Extending spray from three to 30 seconds has a larger effect than extending the wait equally (Chinese 1.57~pp, English 0.39~pp).

\paragraph{Volume labels change while water levels barely move.}
Base FLUX renders the correct 9 and 81~mL labels in 15 and 17 of 32 edits, respectively. Coverage changes by 0.018~pp for a 72~pp target change. OmniGen agrees within 0.39~pp on equivalent 30\% fills, yet 51 of 64 edits fill less than 1\%. For a $60^\circ$ target increase on six dials, base FLUX moves $4.41^\circ$ under direct angles, $5.59^\circ$ under timed rotation and $-1.83^\circ$ under counted steps. On eight held-out canvases, first-step rank-eight adaptation raises realized dose response from 0.99\% to 9.02\% of the target increase and MAE from 16.78 to 19.66~pp.

\section{Depicted area changes little across doses}\label{app:dose}
\begin{figure}[!htbp]
\centering
\includegraphics[width=\textwidth]{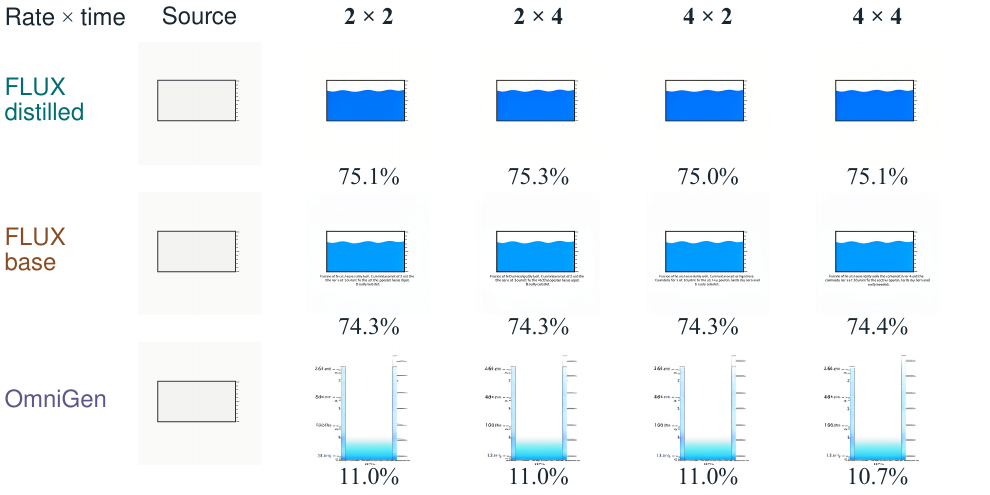}
\caption{\textbf{Dose and water level on a common source tank.} Targets in the 36-unit tank are 11.1\%, 22.2\%, 22.2\% and 44.4\% fill. Labels give blue coverage inside the source frame.}
\label{fig:tank}
\end{figure}

Quadrupling dose changes mean coverage by at most 1.43~pp (Fig.~\ref{fig:tank}), or 1.40~pp for the largest blue component overlapping the cross or tank bottom.
\clearpage
\section{Early context affects coverage through multiple denoiser groups in distilled FLUX}
\label{app:blocks}
\begin{figure}[!htbp]
\centering
\includegraphics[width=\textwidth]{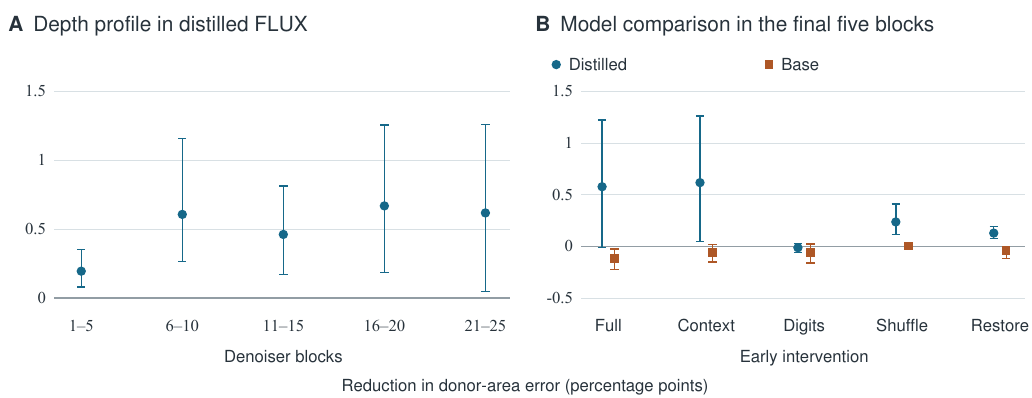}
\caption{\textbf{Coverage transfer across denoiser depth.} (A) Reduction in donor-area error by depth in distilled FLUX. (B) Both models on the same eight-unit layouts. Full exchanges context across all blocks. Context and Digits denote image updates in the final five blocks. Shuffle permutes the context update spatially. Restore retains the group's native update during full exchange. Bars give 95\% paired-layout bootstrap intervals around the means.}
\label{fig:blocks}
\end{figure}

Figure~\ref{fig:blocks}A compares five consecutive five-block groups on eight paired layouts. Each exchange targets the group's change in generated-image features at a common recipient latent, preserving recipient text and reference outputs.

In distilled FLUX, blocks 16--20 and 21--25 reduce donor error by 0.67 and 0.62~pp (final minus preceding group $-0.05$~pp, 95\% CI $-0.18$ to 0.08). Restoring the native update costs 0.45~pp under full exchange (Fig.~\ref{fig:blocks}B, Table~\ref{tab:block-tests}).

\begin{table}[!htbp]
\centering
\caption{Early context in blocks 21--25 of distilled FLUX. Mean donor-error reductions (pp) have paired-layout intervals and Holm-adjusted exact $p$ values. Context and digit norms match.}
\label{tab:block-tests}
\small
\begin{tabular*}{\linewidth}{@{\extracolsep{\fill}}lrrr@{}}
\toprule
Control & Gain & 95\% CI & Adj.\ $p$\\
\midrule
Native & 0.618 & 0.047 to 1.263 & 0.328\\
Digits & 0.631 & 0.063 to 1.283 & 0.328\\
Spatial shuffle & 0.381 & $-0.098$ to 0.880 & 0.328\\
\bottomrule
\end{tabular*}
\end{table}

The final group raises base-FLUX error from 0.77 to 0.83~pp. The paired distilled-minus-base benefit is 0.68~pp (95\% CI 0.11 to 1.30, exact $p=0.086$).

\end{document}